# The discovery of PSR J1833−1034: the pulsar associated with the supernova remnant G21·5−0·9

Y. Gupta[1,*], D. Mitra[1], D. A. Green[2] and A. Acharyya[1,3]

[1] National Centre for Radio Astrophysics, TIFR, Pune University Campus, Pune 411007, India
[2] Mullard Radio Astronomy Observatory, Cavendish Laboratory, Madingley Road, Cambridge CB3 0HE, United Kingdom
[2] National Institute of Technology, Durgapur 713 209, India

**Abstract.** We report the discovery of a young pulsar associated with the supernova remnant G21·5−0·9, using the Giant Metrewave Radio Telescope (GMRT) located near Pune, India. Discovered at a frequency of 610 MHz, PSR J1833−1034 has a period of 61.86 ms and a period derivative of $2.0 \times 10^{-13}$, making it similar to other known young pulsars. The characteristic age of the pulsar is $\approx 4900$ yr, somewhat higher than estimates for the age of the remnant, but not incompatible with it. The pulsar has a spin-down luminosity of $3.3 \times 10^{37}$ erg s$^{-1}$, which is the second highest amongst all the known Galactic pulsars.

## 1 Introduction

Since pulsars are neutron stars that are thought to be produced predominantly from the collapsed stellar cores left behind in supernova explosions, it is natural to expect supernova remnants (SNRs) to harbour young pulsars. Current catalogues[1] of pulsars (see also http://www.atnf.csiro.au/research/pulsar/psrcat/) contain about 1500 Galactic pulsars, but to date only about 20 of these have been clearly associated with one of the 231 known SNRs in our Galaxy[2] (see also http://www.mrao.cam.ac.uk/surveys/snrs/). There are still several well known SNRs for which there are no associated pulsars known. Improving the statistics of the pulsar–SNR associations has several implications, ranging from better understanding of issues related to the evolution of massive stars, the birth rates and evolutions of neutron stars, to providing insights into the evolution of pulsar powered SNRs.

G21·5−0·9 is a SNR that is a strong candidate for harbouring a pulsar. It was first noted[3] as a possible SNR in 1970, when it was proposed that this is one of a small number of remnants that are similar to the Crab Nebula, i.e. a "filled-centre" remnant, with a centrally brightened structure, with flat spectrum synchrotron emission at radio wavelengths. The filled-centre SNR identification for G21·5−0·9 was confirmed[4,5] by subsequent radio observations, that detected its radio emission as clearly polarised, and also by the detection[6] of X-rays from the remnant. These observations revealed centrally brightened emission, approximately circular, about 1 arcmin in extent.

The majority of known Galactic SNRs are of the "shell" type, showing a more or less complete limb-brightened ring of radio emission. Only a small proportion, $\approx 4\%$, are classified as filled-centre remnants, which instead show a centrally brightened structure, presumably due to the injection of energy into the remnant by a central, compact source. In the case of the Crab Nebula – which is the remnant of the SN of AD 1054, and is by far the brightest and best studied filled-centre remnant – it is well known that this remnant is powered by a central pulsar. More recently, however, the improved sensitivity of the *CHANDRA* X-ray satellite led to the detection[7] of a faint outer halo of emission from G21·5−0·9 about 4 arcmin in extent, which surrounds the previously detected centrally brightened core. This outer emission has not to date been clearly detected at radio wavelengths, although its presence is suggested by some observations[8]. Consequently G21·5−0·9 is now classified as a "composite" remnant, i.e. one which contains both the centrally brightened core similar to a filled-centre remnant, and the more extended shell of emission. Currently $\approx 12\%$ of catalogued remnants are classified as composite remnants.

The distance to G21·5−0·9 can be estimated from neutral hydrogen (H I) absorption[9], giving a value of $\approx 4.6$ kpc – using a standard flat Galactic rotation curve model – which is consistent with that deduced from the column density inferred from X-ray observations[10]. At this distance the observed angular size of the faint outer emission of 4 arcmin corresponds to a physical size of 5.2 pc, which is relatively small for a Galactic SNR, implying this is a young remnant, as does its reasonably circular symmetry. Simplistically, taking a constant expansion speed of $\sim 10,000$ km s$^{-1}$, then the age of the remnant would be only $\sim 250$ years; whereas for a constant expansion speed of $\sim 2,000$ km s$^{-1}$ (i.e. comparable to that observed in the

*For correspondence. (e-mail: ygupta@ncra.tifr.res.in)

Crab Nebula) its age would be ∼ 1250 years (see also more sophisticated discussion [11], which leads to an age estimate of between 200 and 1000 years for G21·5−0·9).

Thus, G21·5−0·9 is clearly a good target for looking for a young pulsar counterpart. However, searches for this in both radio [12,13] and X-rays [7,14] have so far been unsuccessful. At radio wavelength the upper limit was 2 to 4 mJy at 400 MHz (assuming a pulse duty cycle of 5%), and at X-rays, the limit is 7.5% to 40% for the pulsed fraction of the emission. Here we report the discovery, using sensitive new observations with the GMRT, of a young pulsar which is the long sought for neutron star associated with G21·5−0·9.

## 2   Observations, Data Analysis and Results

As part of an ongoing search for pulsars in a sample of Galactic SNRs, G21·5−0·9 was observed with the GMRT [15] at 610 MHz on 19 March 2005, for a duration of 2.5 hrs. The observations used 13 of the 14 central square GMRT antennas in a phased array mode [16]. The phased array beam of the central square antennas at 610 MHz is ≈ 2 arcmin, which is well matched to cover the central core of G21·5−0·9. The observations were centred at $18^h 33^m 34^s$, $-10°34'00''$ (J2000.0). The array was phased (once, at the beginning of the observations) using the nearby calibrator source 1822−096. This source was also used to calibrate the flux density scale for the observations, from on-off deflection measurements taken immediately after the observations of the target. For calibration purposes we have taken the flux density of 1822−096 at 610 MHz to be 9.1 Jy, from an interpolation of the 90-cm and 20-cm flux density values available in the VLA calibrator manual (see http://www.aoc.nrao.edu/~gtaylor/calib.html). The total observing bandwidth was 32 MHz centred around 610 MHz, split into two 16 MHz sidebands. The voltage signals from the 256 spectral channels per sideband per polarisation from each antenna were added together in the GMRT Array Combiner to obtain the phased array voltage sum, and then processed in the pulsar back-end to generate 16-bit total intensity data, sampled once per 0.512 ms for each of 512 spectral channels. These data were acquired by a PCI-based data acquisition card, processed in real-time on a computer to check for basic data quality and time fidelity, and then recorded to disk for off-line analysis. These observations had a theoretical 10-sigma detection sensitivity limit of 0.2 mJy at 610 MHz, which is 5 to 10 times better than that of the previous radio searches [12,13] of G21·5−0·9, assuming a median pulsar spectral index $\alpha$ of −1.6 (here $\alpha$ is defined in the sense that flux density $S$ scales with frequency $\nu$ as $\propto \nu^\alpha$).

The recorded data were searched for the presence of dispersed, periodic pulsar signals using the PRESTO software [17], adapted to handle dual sideband GMRT data. The data were dedispersed for a range of trial dispersion measures (DMs) from 0 to 400 pc cm$^{-3}$, in steps of 1 pc cm$^{-3}$. For each dedispersed time series (of 16 million data points), a $2^{24}$ point FFT was performed and the resulting spectra were searched for the presence of period signals using the harmonic summing technique. The analysis of the data for DMs around 170 pc cm$^{-3}$ showed the clear presence of a periodic signal of 61.86 ms, with at least 6 harmonics showing up well above the noise power in the spectrum. The resulting folded profile for the candidate pulsar is shown in Fig. 1, and has an effective signal-to-noise ratio of ∼ 45. The best barycentric period $P$ for this epoch was 61.8639867(92) ms. The pulse profile is quite sharp and narrow, with a half width ≈ 0.065 of the period. Using the flux density calibration scale, we estimate a value of $S_{610\ \text{MHz}} = 0.9 \pm 0.2$ mJy for the flux density of the pulsar at this epoch, where the errors are dominated by the flux density calibration process.

Follow up observations of G21·5−0·9 to confirm the candidate pulsar were carried out at the GMRT on 29 July 2005 (centred at 1068 MHz) and on 30 July 2005 (centred at 618 MHz). For both observations, only one sideband of 16 MHz bandwidth was available for recording. Twelve antennas of the central square were again used in phased array mode, for a duration of 2 hours on each day. Once again 1822−096 was used as the calibration source for phasing the array, and also for flux density calibration. Though no evidence was found for the pulsar in the 1068 MHz data of 29 July 2005, the analysis of the 618 MHz data of 30 July 2005 clearly revealed the presence of the pulsar, albeit with a reduced signal to noise ratio. This detection confirmed the genuine nature of the pulsar. The barycentric period at this epoch was found to be slightly longer, 61.8663333(65) ms, indicating a significant spin-down of the pulsar. We obtained a flux density estimate of $S_{610\ \text{MHz}} = 0.4 \pm 0.1$ mJy at this epoch.

## 3   Interpretation

From our observations, we have enough information to derive several parameters for the newly discovered pulsar, PSR J1833−1034. From the measured change in the barycentric period between the two epochs of observations at 610 MHz, we calculate a period derivative of $\dot{P} = 2.0 \times 10^{-13}$, a value typical for young pulsars in SNRs. From the values of $P$ and $\dot{P}$, we find a characteristic age of $\tau = (P/2\dot{P}) \approx 4900$ yr, a surface magnetic field of $B = 3.2 \times 10^{19} \sqrt{P\dot{P}} \approx 3.6 \times 10^{12}$ G, and a spin-down luminosity of $\dot{E} = 4\pi^2 I \dot{P}/P^3 \approx 3.3 \times 10^{37}$ erg s$^{-1}$ (assuming a typical value of the neutron star's moment of inertia of $I = 10^{45}$ g cm$^2$). From the two epochs of detection, the mean flux density at 610 MHz is 0.65 mJy, and from the non-detection at 1068 MHz, we estimate an upper limit of the pulsar's flux density of $S_{1068 \text{ MHz}} < 0.15$ mJy. This results in a limit on the spectral index of $\alpha < -2.7$, i.e. the pulsar is likely to have a rather steep spectrum. It is possible that day to day variability of the pulsar signal strength makes these flux density and spectral index estimates somewhat tentative – these values will get refined with further observations.

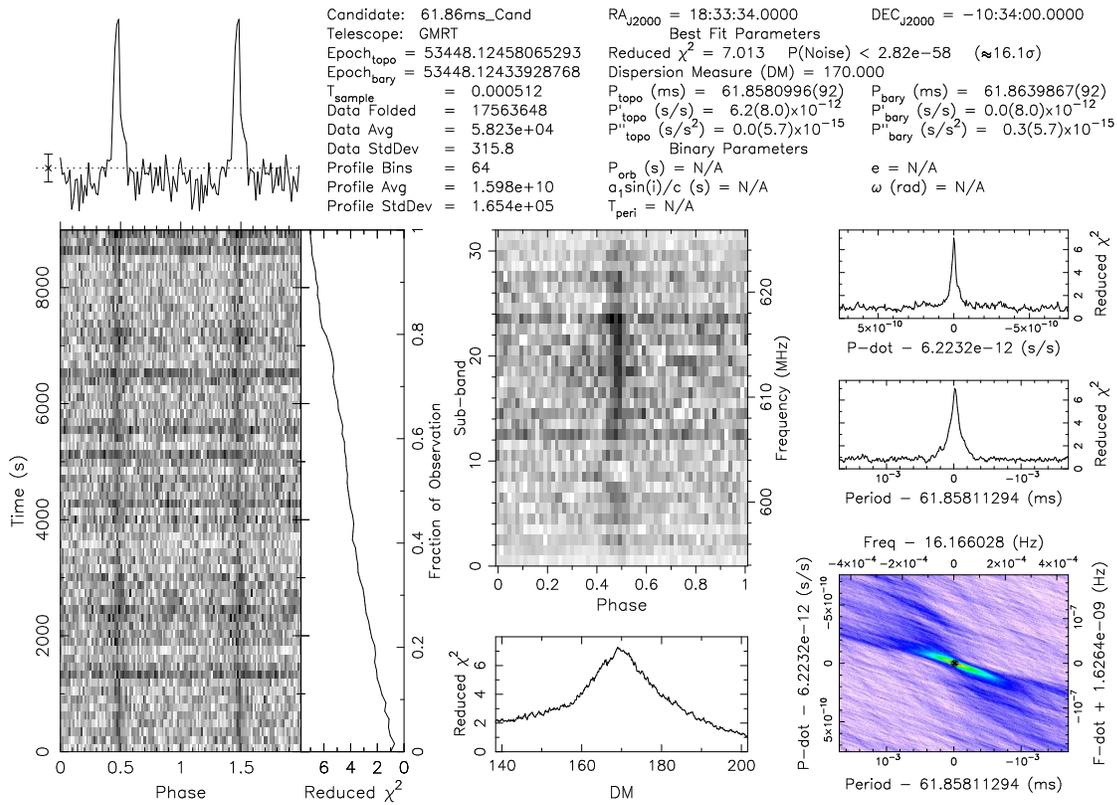

Figure 1: Search analysis results for the detection of PSR J1833−1034. The top left panel shows two pulses of the final folded profile obtained for the best pulsar candidate. The strength of the detection over the 2.5 hr observing duration is fairly constant (left panel), as is its presence across most of the 32 MHz observing band (centre panel). The bottom middle panel shows that the best signal is obtained around a DM of 170 pc cm$^{-3}$ and the panels on the right show the localisation of the pulsar signal in the period and period derivative domains. The best barycentric and topocentric periods for the discovery epoch are also printed.

Table 1: Properties of selected known young Galactic pulsars and the SNRs that contain them.

| pulsar | | | | | | supernova remnant | | | | |
|---|---|---|---|---|---|---|---|---|---|---|
| name | period | period derivative | characteristic age | magnetic field | spin-down luminosity | name | angular size | distance | physical size | date of SN |
| | /ms | /s s$^{-1}$ | /kyr | /G | /erg s$^{-1}$ | | /arcmin | /kpc | /pc | |
| B0531+21 | 33.1 | $4.2 \times 10^{-13}$ | 1.2 | $3.9 \times 10^{12}$ | $4.6 \times 10^{38}$ | Crab Nebula | $5 \times 7$ | 1.9 | $2.8 \times 3.9$ | AD 1054 |
| J1833−1034 | 61.9 | $2.0 \times 10^{-13}$ | 4.9 | $3.6 \times 10^{12}$ | $3.3 \times 10^{37}$ | G21·5−0·9 | 4 | 4.6 | 5.4 | – |
| J1811−1925 | 64.7 | $0.4 \times 10^{-13}$ | 23 | $1.8 \times 10^{12}$ | $0.6 \times 10^{37}$ | G11·2−0·3 | 4 | 4.4 | 5.1 | AD 386? |
| J0205+6449 | 65.7 | $1.9 \times 10^{-13}$ | 5.4 | $3.7 \times 10^{12}$ | $2.6 \times 10^{37}$ | 3C58 | $5 \times 9$ | 3.2 | $2.8 \times 5.0$ | AD 1181 |

The observed and derived parameters for PSR J1833−1034 are quite typical of known young pulsars seen in supernova remnants, some of which are summarised in Table 1. The properties of pulsars – apart from for PSR J1833−1034 – are from version 1.23 of PSRCAT[1] (see also http://www.atnf.csiro.au/

research/pulsar/psrcat/). The properties of the SNRs are from the 2004 January catalogue of Galactic SNRs[2] (see also http://www.mrao.cam.ac.uk/surveys/snrs/). In the cases of the Crab Nebula and 3C58 the age of the SN that produced the remnants and the pulsars is well established, whereas in the case of G11·2−0·3 the association with an historical SN is not certain[18]. In many characteristics, this new pulsar is very similar to PSR J0205+6449, the pulsar in the SNR 3C58[19]. It is notable that the inferred spin-down luminosity for PSR J1833−1034 makes it the second most energetic amongst the Galactic pulsars (behind only the Crab), and the fourth highest amongst all pulsars, including those in the Magellanic Clouds. The period of the pulsar is in fact similar to that predicted[10] using models for the pulsar wind[20] – for a spin-down energy loss rate of $3.3 \times 10^{37}$ erg s$^{-1}$, and an age of $\approx 1000$ yr, the predicted period is $\sim 75$ ms, close to that observed. The characteristic age of $\approx 4900$ yr puts PSR J1833−1034 amongst the ten youngest pulsars in the Galaxy. Though this characteristic age is somewhat larger than typical estimates of the age of G21·5−0·9 (as described in Section 1), the discrepancy is not serious, given the common understanding that pulsar characteristic ages tend to overestimate the true age of young pulsars, when the pulsar's present period need not be much slower than that at birth[21]. To a lesser extent, a braking index less than the canonical value of 3.0[22] can also produce such a difference.

The DM value of $170.0 \pm 1.0$ pc cm$^{-3}$ indicates that PSR J1833−1034 is a relatively nearby object. In this direction of the Galaxy, existing electron density models[23,24] predict a distance in the range of 3.3 to 3.7 kpc for this DM which are somewhat smaller than distance of 4.6 kpc inferred for the SNR from H I measurements (see Section 1). However, given the uncertainties involved ($\sim 25\%$ in the electron density models and at least $\sim 0.5$ kpc in the H I distance estimates), the difference is not significant, thereby further supporting an association between the two. Using an intermediate distance of $d \approx 4$ kpc, the spectral radio luminosity of PSR J1833−1034 at 610 MHz is $L_{610\ \mathrm{MHz}} \approx 10$ mJy kpc$^2$. Using our inferred spectral index limit, this translates to $L_{1400\ \mathrm{MHz}} < 1.1$ mJy kpc$^2$, making this one of the least luminous of the known young radio pulsars in a SNR.

Regular follow-up observations of PSR J1833−1034 are now underway at the GMRT. The timing measurements will result in better estimates for several of the pulsar parameters. The position of the pulsar, which is presently known only to the accuracy of the beam-width of the phased array, i.e. $\approx 2$ arcmin, will also become known more precisely. More reliable estimates of the pulsar's flux density and spectral index will also be obtained.

## 4 Summary

As part of an ongoing survey of Galactic supernova remnants, sensitive observations made with the GMRT at 610 MHz have detected the young, energetic pulsar PSR J1833−1034. With measured period and period derivative of 61.86 ms and $2.0 \times 10^{-13}$, respectively, this pulsar ranks amongst the 10 youngest pulsars in the Galaxy and also has the second highest spin-down energy loss rate of all pulsars in the Galaxy. These properties of a young pulsar, along with the positional coincidence with the compact SNR G21·5−0·9, imply that PSR J1833−1034 is the neutron star born during the supernova explosion that formed this remnant. This association is supported by the fact that the characteristic age and dispersion distance of PSR J1833−1034 are in reasonable agreement with estimates of the age and distance of G21·5−0·9. This discovery – the first of a pulsar in a SNR by the GMRT – reveals the potential of the GMRT to discover several more of such faint, elusive pulsars in supernova remnants.

## 5 Acknowledgements

We thank the staff of the GMRT for help with the observations. The GMRT is run by the National Centre for Radio Astrophysics of the Tata Institute of Fundamental Research. We also thank S. Sarala for help in setting up the search analysis software at NCRA, and R. Nityananda for useful comments on the manuscript.

*Note added in proof:* After the acceptance of this paper for publication in *Current Science*, we have learnt (S.Ransom, private commun.) that the pulsar J1833-1034 has now been detected in observations at the Parkes and Greenbank Radio Telescopes.